\documentclass{PoS}

\usepackage{amsmath,bbm,slashed,booktabs}

\usepackage{ragged2e}

\newcommand{\eg}{\emph{e.g.}~}
\newcommand{\eq}[1]{\begin{equation} #1 \end{equation}}
\newcommand{\eqa}[1]{\begin{eqnarray} #1 \end{eqnarray}}

\usepackage{calrsfs}
\DeclareMathAlphabet{\pazocal}{OMS}{zplm}{m}{n}

\title{Charmless Non-Leptonic Multi-Body B decays}

\ShortTitle{Charmless Non-Leptonic B decays}

\author{\speaker{Javier Virto}\thanks{Preprint number: NIOBE-2016-01}
\vspace{2mm}\\
Albert Einstein Center for Fundamental Physics, Institute for Theoretical Physics,\\
University of Bern, CH-3012 Bern, Switzerland.
\vspace{2mm}\\
E-mail: \email{jvirto@itp.unibe.ch}}

\abstract{
I review the theoretical and phenomenological status of two- and three-body charmless non-leptonic $B$ decays,
with an emphasis on factorization approaches. Most of the material presented here is based on talks
given by the participants of the workshop ``Future Challenges in Non-Leptonic $B$ Decays: Theory and Experiment'',
held at Bad Honnef in February 2016.}

\FullConference{
Flavor Physics and CP Violation,\\
6-9 June 2016\\
Caltech, Pasadena CA, USA}

\begin{document}

\section{Introduction and Motivation}

\subsection{Definitions}

Non-leptonic $B$ decays are exclusive decays of the form $B\to h_1 \cdots h_n$ with $h_i$ any heavy or light
hadrons. Charmless non-leptonic decays are non-leptonic decays with no charmed hadrons in the final state
(and excluding $c\bar c$ states). We will mostly focus on charmless decays to two ($B\to M_1M_2$) and
three ($B\to M_1M_2M_3$) final mesons, but most generalities are common to all non-leptonic modes.

\subsection{Non-leptonic B decays within the global arena of particle physics}

There are a number of open issues in our understanding of the physics of the elementary particles and their
interactions which our current theory --the Standard Model (SM)-- does not seem to be able to answer. These
questions are related to gauge symmetry, electroweak symmetry breaking, flavor and CP (including baryogenesis),
astrophysics and cosmology (dark matter, dark energy and inflation), and gravity. Many extensions of the SM
addressing some of these issues have been put forward, and many are perfectly plausible given our current
theoretical knowledge and experimental record. While it is possible that purely theoretical work may narrow down
in the future the number of viable models, it is clear that the fast track is to obtain hints from experiment.
A more complete theory that addresses all or some of these issues, while at the same time sharing the many
outstanding successes of the SM, will very likely as well modify the predictions for current and future
laboratory experiments which study collisions and decays of known particles. Establishing such deviations
with respect to SM expectations will not only provide \emph{direct} evidence for the need of a non-trivial
extension of the SM, but also specific hints of what this extension should look like. This is arguably the
most important task in particle physics today.

Testing the SM requires first to measure its free parameters precisely and to understand how to make precision
calculations. Most of the free parameters of the SM are related to flavor, such as the entries of the CKM
matrix --which govern the physics of flavor in the quark sector (flavor transitions of hadrons). Non-leptonic
$B$ decays are an essential input in CKM fits, and necessary for the direct measurement of the CKM angles
$\alpha$, $\beta$ and $\gamma$ (see \eg\cite{1501.05013}), thus providing, in addition, valuable
tests of the SM mechanism of CP violation. They also provide direct access to
the study of $B_q-\bar B_q$ mixing ($\Delta B=2$ transitions) through the interference of CP-conjugated
decays into final CP eigenstates. From the huge number of different non-leptonic $B$ decays accessible
experimentally, some are mediated at tree level in the SM, while some arise only at the loop level; some
are dominated by a single SM amplitude, while some are the result of interference of two amplitudes of similar
size with different weak and strong phases. This results in very broad phenomenological applications from
SM studies to New Physics (NP) searches, and including hadronic physics.

\subsection{Non-leptonic $B$ decays in the context of strong interactions}

Any process involving hadrons is probing the physics of QCD bound states in some way. Therefore one is
forced to either make full computations in a non-perturbative regime, or to isolate the contributions
sensitive to infrared (IR) physics, parametrize them by a few ``universal'' quantities, and subsequently
(a) calculate them, (b) extract them from experiment, or (c) build observables where these cancel out.
So far, we can only calculate non-perturbatively a few simple objects such as decay constants
(matrix elements like $\langle 0 | \bar d_L \gamma^\mu b_L | \bar B\rangle$) or form factors
(matrix elements such as $\langle \pi | \bar d_L \gamma^\mu b_L | \bar B\rangle$). These calculations are
based on numerical simulations (in the framework of Lattice QCD, see \eg\cite{1607.00299}),
or on operator product expansions and dispersion relations (within the framework of QCD sum rules, see
\eg\cite{0010175}). Decay constants and form factors are enough for predictions of leptonic
(\eg $B^-\to \ell^- \bar \nu_\ell$ or $B_s\to \ell^+\ell^-$) and
semileptonic (\eg $B^-\to \pi^0 \ell^- \bar \nu_\ell$) decays --at least to leading order in QED--,
but not for non-leptonic decays such as $B\to \pi\pi$.

Isolating IR effects is a particular case of \emph{scale separation} in quantum field theory, which --if
the scales are widely separated-- is achieved most conveniently in the framework of effective field
theory (EFT). In the case of weak meson decays, a first step is to separate the scale of weak interactions
($\sim M_W$) from the scale of hadronic physics ($m_b$ or lower). This leads to the \emph{Weak Effective
Theory} (see \eg\cite{9512380}) where flavor-changing transitions are mediated by dimension-six operators:
\eq{
{\cal L}_W = {\cal L}_{QCD+QED}^{\rm no\ top} - \frac{4G_F}{\sqrt{2}}
\sum_{p=u,c} \Bigg\{ \lambda_p^{(D)}\ \bigg[ C_1\,Q_1^p + C_2\,Q_2^p +\sum_{i=3\cdots 6,8} C_i\,Q_i \bigg] + h.c. \Bigg\}  + \cdots
}
Here the CKM prefactors $\lambda_p^{(D)} \equiv V_{pb} V^*_{pD}$ ensure the Wilson coefficients
$C_i$ are independent of CKM elements in the SM once CKM unitarity is used, and we have only
written down explicitly the dimension-six operators most relevant for charmless non-leptonic $b\to D\bar q q$ and $\bar b\to \bar D\bar q q$
transitions, with $D=\{d,s\}$ and $q=\{u,d,s\}$. These include current-current operators
$Q_{1,2}^p$, QCD-penguin operators $Q_{3\cdots 6}$ 
and the chromomagnetic operator $Q_8$ (see \eg\cite{9711280}):
\begin{align}
Q_{1,2}^p &= (\bar D_L \gamma^\mu \{T^a,\mathbbm{1}\} p_L) (\bar p_L \gamma_\mu \{T^a,\mathbbm{1}\} b_L) \, , &
Q_{3,4}  &= (\bar D_L \gamma^\mu \{\mathbbm{1},T^a\} b_L) \sum_q (\bar q \gamma_\mu \{\mathbbm{1},T^a\} q) \, ,\nonumber\\
Q_{5,6}  &= (\bar D_L \gamma^\mu\gamma^\nu\gamma^\rho \{\mathbbm{1},T^a\} b_L) \sum_q (\bar q \gamma_\mu\gamma_\nu\gamma_\rho \{\mathbbm{1},T^a\} q) \, , &
Q_{8\;\;\;\,} &= -\displaystyle\frac{g_s}{16\pi^2} \, m_b \; \bar D_L \; \sigma_{\mu\nu} G^{\mu\nu} b_R \; , \quad \nonumber
\end{align}
and its hermitian conjugates.
Electroweak penguin operators $\sim \sum_q e_q (\bar D\Gamma b)(\bar q \Gamma' q)$ can also be included, but
their Wilson coefficients are small in the SM.
Other operators with negligible SM Wilson coefficients include scalar operators $\sim (\bar s_L b_R)(\bar q_L q_R)$,
or operators with opposite chirality (\eg$Q_8' \sim \bar D_R \sigma^{\mu\nu} G_{\mu\nu} b_L$).
All of them are potentially relevant beyond the SM.
Additional ``evanescent'' operators are needed for renormalization in dimensional regularization at higher
orders in QCD (see \eg\cite{0411071}).  

Non-leptonic $B$-decay amplitudes are then given by:
\eq{
A(\bar B \to f) =  \lambda_u^{(D)} (T^u_f + P_f) + \lambda_c^{(D)} (T^c_f + P_f)
\label{amp}
}
with
\eq{
T^p_f = -\frac{4 G_F}{\sqrt{2}} \sum_{i=1,2} C_i(\mu)\langle f |Q_i^p(\mu)|\bar B\rangle\ ,\quad
P_f = -\frac{4 G_F}{\sqrt{2}} \sum_{i=3\cdots 6,8} C_i(\mu)\langle f |Q_i(\mu)|\bar B\rangle\ .
}
Note that in the case of charmless decays $T^c_f$ is purely the result of a penguin contraction (and thus ``$T$''
does not necessarily mean ``Tree'').
For $\mu\sim m_b$, the matrix elements of the operators
do not depend on any scale larger than $m_b$ (all the dependence on the weak and, possibly, NP scales is
contained in $C_i(\mu)$). At the same time the Wilson coefficients do not depend on any IR scale and are thus
perturbatively calculable.
The values of the Wilson coefficients $C_i(\mu)$ in the SM at the
renormalization scale $\mu\sim m_b$ can be calculated via the usual matching-and-running procedure, and
are known to next-to-next-to-leading logarithmic (NNLL) accuracy \cite{9910220,0401041,0411071,0504194,0612329},
see Table~\ref{tabWCs}.

\begin{table}
\renewcommand{\arraystretch}{1.2}
\small
\centering
\begin{tabular}{@{}ccccccc@{}c}
\toprule
$C_1(\mu_b)$ &   $C_2(\mu_b)$ &  $C_3(\mu_b)$ &  $C_4(\mu_b)$
& $C_5(\mu_b)$ & $C_6(\mu_b)$ &  $C_8(\mu_b)$ \\[1mm]
\hline
\makebox[15mm]{-0.2632} & \makebox[15mm]{1.0111} & \makebox[15mm]{-0.0055} & \makebox[15mm]{-0.0806} & \makebox[15mm]{0.0004} & \makebox[15mm]{0.0009}  & \makebox[15mm]{-0.1792} \\
\bottomrule
\end{tabular}
\caption{$\overline{\rm MS}$ NNLL Wilson coefficients at the scale $\mu_b=4.8$~GeV.}
\label{tabWCs}
\end{table}

The challenge is to calculate the matrix elements $\langle M_1M_2\cdots |Q_i(\mu)|\bar B\rangle$ in QCD.
This is a very complicated task, not completely understood so far, and which constitutes yet another
strong motivation for the study of non-leptonic $B$ decays: they teach us about QCD. For example, the
soft-collinear effective theory (with a wide range of applications, from heavy-meson decays to collider physics
and gravity, see \eg\cite{1410.1892}), was first developed to describe $B$-meson
decays~\cite{0011336,0109045,0206152,0211358}.

\subsection{Soft-collinear factorization}
\label{scfact}

The matrix elements $\langle M_1M_2 |Q_i(\mu)|\bar B\rangle$ at $\mu\simeq m_b$ depend on three different
momentum scales: (1) a ``hard'' scale $p_h^2\sim m_b^2$ associated to the energy of the process and the choice
of renormalization scale; (2) a ``soft'' scale $p_s^2\sim \Lambda_{\rm QCD}^2$ associated with the dynamics of
light degrees of freedom within the $B$ and light mesons; and (3) a ``hard-collinear'' scale
$p_{hc}^2\sim m_b \Lambda_{\rm QCD}$ associated with a momentum transfer that would give a soft light parton
in the $B$-meson a large energy ($\sim m_b$) and a low virtuality ($\sim \Lambda_{\rm QCD}$), so as to become
part of one of the final mesons. Such a large-energy-low-virtuality momentum is called a ``collinear''
momentum $p_c$ (note that $p_c^2\sim \Lambda_{\rm QCD}$). In the two-body final state there are two different
collinear momenta: $p_c$ and $p_{\bar c}$ in opposite directions, corresponding to partons in $M_1$ and $M_2$.
In the heavy-quark limit these three scales are widely separated: $p_h^2\gg p_{hc}^2\gg p_{s,c,\bar c}^2$,
calling for a scale separation within EFT. Scale separation leads often to useful factorization ``theorems'';
in this case integrating out hard scales at the leading power leads to ``soft-collinear factorization''
(see \eg\cite{1501.07374,bell}) with decoupling of anti-collinear modes.

The first step is to integrate out from QCD the scale $p_h^2$. This leads to an EFT called SCET-1. The
matching condition for a QCD operator $Q_i$ in terms of SCET-1 operators $O^I$, $O^{II}$ is given by:
\eqa{
Q_i &=& \int dt\, \widetilde T^I(t) O_i^I(t) + \int dt\, ds\, \widetilde T^{II}(t,s) O_i^{II}(t,s)\\[2mm]
O_i^I(t) &=&
\big[ (\bar\chi W_{\bar c})(t n_-)\Gamma_i^1(W_{\bar c}^\dagger \chi)(0) \big]
\big[ (\bar\xi W_c)(0) \Gamma_i^2 h_v(0)\big]\\[2mm]
O_i^{II}(t,s) &=&
\big[ (\bar\chi W_{\bar c})(t n_-)\Gamma_i^3(W_{\bar c}^\dagger \chi)(0) \big]
\big[ (\bar\xi W_c)(0) \Gamma_i^4 (W_c^\dagger i \hspace{1mm}/\hspace{-2.5mm} D_{\bot c} W_c)(s n_+)
\Gamma_i^5 h_v(0) \big]
}
where $\xi$, $\chi$ and $h_v$ are collinear, anti-collinear and heavy quark fields,
$W_{c,\bar c}$ are collinear and anticollinear Wilson lines,
$n_\pm$ are light-cone vectors in the collinear and anti-collinear directions,
and $\Gamma_i^j$ are Lorentz structures. The functions $\widetilde T^{I,II}$ are perturbative Wilson coefficients
that depend only on the hard scale.
In SCET-1 there are no leading power interactions between
anti-collinear and soft or collinear modes, and the anti-collinear sector decouples. Thus the matrix elements
of $Q_i$ are proportional to a light-cone distribution amplitude (LCDA) of a light meson:
\eq{
\langle M_2 |(\bar\chi W_{\bar c})(t n_-)\Gamma_i(W_{\bar c}^\dagger \chi)(0)|0\rangle \sim \phi_{M_2}(t)\ ,
}
where it is assumed that $M_2$ has anti-collinear momentum. The matrix elements of heavy-collinear currents
between $B$ and $M_1$ still depend on the hard-collinear scale. Hard-collinear modes are integrated out
at a second step, leading to an EFT called SCET-2, containing only soft and (anti)collinear modes.
Hard-collinear factorization works for $O^{II}$, leading to:
\eq{
\langle M_1 | (\bar\xi W_c)(0) \Gamma_i
(W_c^\dagger i \hspace{1mm}/\hspace{-2.5mm} D_{\bot c} W_c)(s n_+) \Gamma'_i\,h_v(0)|\bar B \rangle
\sim \int d\omega\, du\, J_i(s,w,u) \phi_B(\omega) \phi_{M_1}(u)
}
where $J_i(s,\omega,u)$ is a hard-collinear matching\ coefficient, which is perturbative provided the
hard-collinear scale $\sqrt{m_b \Lambda_{\rm QCD}}$ is perturbative. Hard-collinear factorization
fails for $O^{I}$, so that the form factor $\langle M_1 | (\bar\xi W_c)(0) \Gamma\, h_v(0)|\bar B \rangle$,
which depends on soft and hard-collinear momenta, cannot be factorized. This is a long-standing
problem~\cite{0211069,0311335,0311345}. In practice, this is part of the full QCD $B\to M_1$ form factor $F^{BM_1}$, which
appears in factorization formulas.

In the case matrix elements with more than two final-state particles
(\eg$\langle M_1M_2M_3 |Q_i(\mu)|\bar B\rangle$), the identification of the relevant
scales is much less straightforward. This depends on the specific \emph{kinematics}
of the decay. This discussion is relegated to Section~\ref{three-body}.

\section{Two-body decays}

\subsection{Factorization formula for two-body decays at the leading power}

The arguments laid down in Section~\ref{scfact} lead to a factorization formula for charmless two-body
$B$ decays at the leading power in $\Lambda_{\rm QCD}/m_b$, first put forward in~\cite{9905312,0006124}.
It should be remarked that after 20 years of intense research these papers are not outdated in any way
and remain state-of-the-art: much has been understood conceptually since then but the formulation has not
changed a bit. In essence, the matrix element of an operator $Q_i$ is given by:
\eq{
\langle M_1 M_2 |Q_i|\bar B\rangle = F^{BM_1} \int du\,  T^I_i(u) \phi_{M_2}(u) +
\int d\omega\,du\,dv\,  T^{II}_i(\omega,u,v) \phi_B(\omega)\phi_{M_1}(u)\phi_{M_2}(v),
\label{qcdf}
}
where $F^{BM}$ is a form factor in QCD, $\phi_M$ are LCDAs of light and heavy mesons, and the
(perturbative) ``hard-scattering kernels'' $T^{I,II}_i$ are related to the SCET matching coefficients $\widetilde T^{I,II}$
and $J_i$ in Section~\ref{scfact}. The notation is such that $M_1$ picks the $B$-meson spectator quark;
if $M_2$ can also pick the spectator, there is an additional corresponding term proportional to $F^{BM_2}$.
$T^I(u)=1+{\cal O}(\alpha_s)$ arises from vertex corrections already at the leading order, while
$T^{II}(\omega,u,v)={\cal O}(\alpha_s)$ starts at next-to-leading order and involves spectator scattering,
and it is power suppressed if $M_1$ is heavy (not in charmless decays).

The factorization formula~(\ref{qcdf}) is valid only up to ${\cal O}(\Lambda/m_b)$ corrections but 
(presumably) to all orders in $\alpha_s$.
Formally, this has been proven explicitly up to NNLO. Assuming that the SCET contains all
the relevant IR degrees of freedom (which is the standard assumption), leads to an all-order proof.

\subsection{Status of perturbative calculations}

The original papers on QCDF (\eg\cite{0308039}) contain already all next-to-leading order (NLO) corrections
(\emph{i.e.}~${\cal O}(\alpha_s)$) to the hard-scattering kernels $T^I$, $T^{II}$ for both tree and penguin
topologies. The calculation of next-to-next-to-leading order corrections (\emph{i.e.}~${\cal O}(\alpha_s^2)$)
is a much more demanding task, which has been almost completed during the last decade. These include:
two-loop vertex corrections~\cite{0705.3127,0902.1915,0911.3655} and two-loop penguin and
one-loop chromomagnetic operator contributions~\cite{1107.1601,1507.03700} to $T^I$,
as well as one-loop vertex corrections~\cite{0512351,0608291,0709.3214}
and one-loop penguin contributions~\cite{0610322,0706.3399} to $T^{II}$.
This is summarized in Table~\ref{tablepc}.
NNLO penguin contributions to $T^I$ from $Q_{1,2}^c$ are particularly difficult as they require the
evaluation of a large number of two-loop Feynman integrals with three scales ($m_b$,$m_c$, $um_b$),
with a non-trivial threshold at $(1-u) m_b^2 \sim 4 m_c^ 2$~\cite{1410.2804}.  
Missing NNLO pieces include two-loop vertex and penguin corrections from the penguin
operators~$Q_{3\cdots 6}$, which are nevertheless numerically subleading for tree decays.

\begin{table}
\centering
\includegraphics[width=14cm]{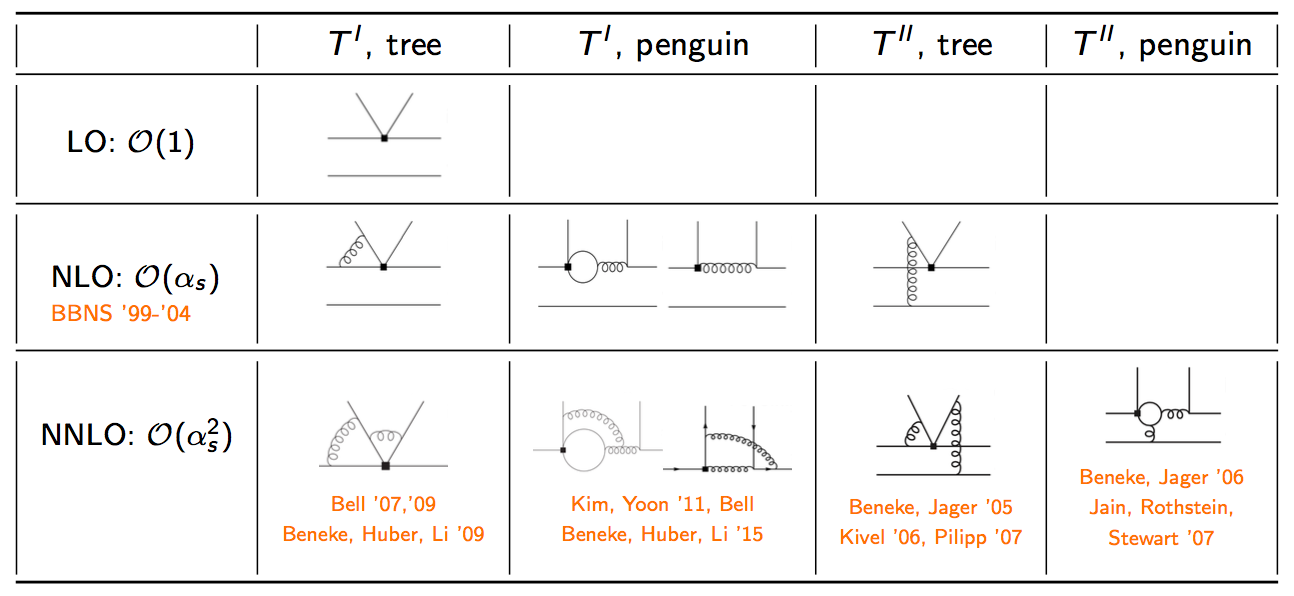} 
\caption{Summary of the status of perturbative calculations of charmless
two-body $B$ decays in QCDF.}
\label{tablepc}
\end{table}

At leading power, strong phases appear first at NLO. Therefore the first correction to CP asymmetries
comes from NNLO corrections. This is the main motivation for the NNLO calculation.
We now summarize briefly the phenomenology~\cite{martin}.

\subsection{Tree-dominated decays}

Tree-dominated decays are those receiving CKM-allowed vertex contributions from current-current operators.
We take as an example $B\to\pi\pi$. In this case the amplitudes are given by:
\eqa{
\sqrt{2}\,A(B^-\to \pi^-\pi^0) &=& \lambda_u^{(d)}\ \big[a_1(\pi\pi) + a_2(\pi\pi)\big]\ A_{\pi\pi}\\[2mm]
-A(\bar B^0\to \pi^0\pi^0) &=& \lambda_u^{(d)}\ \big[a_2(\pi\pi) - \hat \alpha_4^u(\pi\pi)\big]\ A_{\pi\pi}
- \lambda_c^{(d)}\ \hat \alpha_4^c(\pi\pi) \ A_{\pi\pi}
}
and $A(\bar B^0\to \pi^+\pi^-) = A(\bar B^0\to \pi^0\pi^0) + \sqrt{2} A(B^-\to \pi^-\pi^0)$. Here we have ignored
contributions from electroweak penguins and annihilation topologies (although
$\hat \alpha_4^p = \alpha_4^p + \beta_3^p$ contains an annihilation contribution $\beta_3^p$).
$a_1$ and $a_2$ are color-allowed and color-suppressed tree amplitudes.
$\alpha_4^p$ contains penguin contractions of current-current operators, and will be considered later.
Since $a_{1,2}\gg \alpha_4$, and $\lambda_u^{(d)}\sim \lambda_c^{(d)}$, tree decays are dominated by
the tree amplitudes $a_{1,2}$.
At NNLO~\cite{0911.3655}:
\eqa{
a_1(\pi\pi) &=& 1.009 + [0.023 + 0.010\,i]_{\rm NLO} + [0.026 + 0.028\,i]_{\rm NNLO}\nonumber\\[2mm]
&& - \bigg[ \frac{r_{\rm sp}}{0.485} \bigg] \Big\{ 0.015+[0.037+0.029\,i]_{\rm NLOsp}+[0.009]_{\rm tw3} \Big\}
= 1.00 + 0.01\,i\ ,
\label{a1}\\[3mm]
a_2(\pi\pi) &=& 0.220 - [0.179 + 0.077\,i]_{\rm NLO} - [0.031 + 0.050\,i]_{\rm NNLO}\nonumber\\[2mm]
&& + \bigg[ \frac{r_{\rm sp}}{0.485} \bigg] \Big\{ 0.123+[0.053+0.054\,i]_{\rm NLOsp}+[0.072]_{\rm tw3} \Big\}
= 0.26 - 0.07\,i\ ,
\label{a2}
}
where $r_{\rm sp} = 9 f_\pi \hat f_B/(m_b f_+^{B\pi}(0)\lambda_B)$ is a normalization related to the
hard-spectator contributions (\emph{i.e.} $T^{II}$), most notably proportional to the inverse moment
$\lambda_B^{-1}$ of the $B$-meson LCDA. The perturbative expansion is seen to be well behaved, taking into
account that the NLO contribution to $a_2$ lifts color suppression, while the opposite is true for $a_1$.
These two amplitudes must be scale-independent, and indeed the $\mu$-dependence stabilizes at NNLO for the
real parts (no so much for the imaginary parts, as the LO contribution is real). Radiative corrections are
relatively large, but significant cancellations occur between the form factor and spectator terms.
The color suppressed amplitude is dominated by the spectator scattering contribution, because the NLO+NNLO
corrections to the form factor term [first line in Eq.~(\ref{a2})] cancel almost completely the LO term. Therefore the
amplitude $\bar B \to\pi^0\pi^0$ has a strong dependence on $\lambda_B$ (one finds $a_2\sim 0.26 \to 0.51$
when $\lambda_B \to \lambda_B/2$).

For $\lambda_B(1~{\rm GeV}) = 0.35 \pm 0.15~{\rm GeV}$, all branching fractions for tree decays ($B\to \pi\pi$, $B\to \pi\rho$,
$B\to \rho\rho$) agree well with experimental measurements within uncertainties, except for
very slight tensions in $\bar B\to \pi^+\pi^-$, $\bar B\to \pi^-\rho^+$ and $\bar B \to \pi^0\rho^0$,
and a significant and persistent tension in $\bar B \to \pi^0\pi^0$:
$10^6 BR(\bar B \to \pi^0\pi^0)_{\rm th} = 0.33^{+0.11+0.42}_{-0.08-0.17}$ vs.
$10^6 BR(\bar B \to \pi^0\pi^0)_{\rm exp} = 1.91\pm 0.23$ (HFAG 2013). It turns out that a lower value for
$\lambda_B(1~{\rm GeV}) \sim 0.20~{\rm GeV}$ improves the agreement of all these modes, and enhances
significantly $10^6 BR(\bar B \to \pi^0\pi^0)_{\rm th}= 0.63^{+0.12+0.64}_{-0.10-0.42}$, bringing it closer to
the experimental average, but still far away. 
Notably, a new Belle analysis~\cite{belleBpipi} reports $10^6 BR(\bar B \to \pi^0\pi^0)_{\rm Belle} = 0.90\pm 0.16$,
and would agree within uncertainties with the theory prediction, assuming such a low value for $\lambda_B$.
Including this new Belle measurement, the HFAG experimental average becomes
$10^6 BR(\bar B \to \pi^0\pi^0)_{\rm exp} = 1.17\pm 0.13$~\cite{HFAG}. This mode is extremely difficult for LHCb.
A precise independent Belle-II measurement will certainly be very welcome.

\subsection{Inverse moment of the $B$-meson LCDA}

The $B$-meson light-cone distribution amplitude $\phi_B^+(\omega,\mu)$ is defined by:
\eq{
i\, f_B\, m_B\, \phi_B^+(\omega,\mu) = \frac1{2\pi} \int dt\,e^{i\omega t}\, \langle 0|\bar q(tn)\,[tn,0]\,\slashed{n}\,
\gamma_5\, h_v(0) |\bar B (m_B v) \rangle\ ,
}
where $n^\mu$ is a light-cone vector with $n\cdot v =1$, and $[tn,0]$ is a straight Wilson line along $n^\mu$.
The parameter $\lambda_B$ is given by the following inverse moment of $\phi_B^+$:
\eq{
\lambda_B^{-1}(\mu) \equiv \int_0^\infty \frac{d\omega}{\omega} \phi_B^+(\omega,\mu)\ .
}
Given the situation with color-suppressed tree-dominated decays,
it is important to determine $\lambda_B$ with high accuracy.
Direct application of QCD sum-rules provides the following estimate~\cite{Braun:2003wx}:
$\lambda_B\equiv \lambda_B(1~\mbox{GeV})= 0.46\pm 0.11~\mbox{GeV}\,$.
Comparing the LCSRs with pion~\cite{Khodjamirian:2011ub} and $B$-meson DAs for the $B\to\pi$ form factor
gives a compatible result~\cite{Wang:2015VGV} $\lambda_B= 0.358^{+0.038}_{-0.030}$~GeV.
A direct experimental extraction is
also possible, from the branching ratio of the radiative leptonic decay $B\to \gamma\ell \nu_\ell$,
which is very sensitive to $\lambda_B$.
Combining the experimental bound from Belle~\cite{Heller:2015vvm}
with the theory prediction~\cite{Beneke:2011nf,Braun:2012kp}
results in the lower limit $\lambda_B> 238$~MeV at $90\%$~C.L., in agreement with the above estimates based on
QCD sum rules.
This limit begins to challenge the lower values around $\lambda_B=200-250 $~MeV preferred by the
QCD factorization analysis of $B\to \pi\pi$. Improved measurements of the $B\to \gamma\ell \nu_\ell$
branching fraction at Belle-II will be essential to resolve this puzzle.
Theoretical improvements in the prediction of $B\to \gamma\ell \nu_\ell$ are also important.
Recent progress includes the calculation of subleading power contributions at one-loop, and
three-particle corrections~\cite{Wang:2016qii}.

\subsection{Penguin-dominated decays}

Penguin-dominated decays are those for which vertex contributions from current-current operators are either absent,
or CKM-suppressed with respect to penguin contractions. Taking as an
example $\bar B \to K\pi$, we have (ignoring electroweak penguins and annihilation topologies):
\begin{alignat}{2}
A(B^-\to \pi^-\bar K^0) &= \lambda_u^{(s)}\ \hat \alpha_4^u(\pi \bar K)\ A_{\pi\bar K}
&&+ \lambda_c^{(s)}\ \hat \alpha_4^c(\pi\bar K) \ A_{\pi\bar K}\\[2mm]
\sqrt{2}\,A(B^-\to \pi^0 K^-) &=
\lambda_u^{(s)} \Big\{ \big[a_1(\pi\bar K) + \hat \alpha_4^u(\pi\bar K)\big] \,A_{\pi\bar K}
+ a_2(\bar K\pi) \, A_{\bar K \pi} \Big\}
&&+ \lambda_c^{(s)}\ \hat \alpha_4^c(\pi\bar K)\ A_{\pi\bar K}
\\[2mm]
A(\bar B^0\to \pi^+ K^-) &= \lambda_u^{(s)}\ \big[a_1(\pi\bar K) + \hat \alpha_4^u(\pi\bar K)\big]\ A_{\pi\bar K}
&&+ \lambda_c^{(s)}\ \hat \alpha_4^c(\pi\bar K) \ A_{\pi\bar K}
\end{alignat}
and $\sqrt{2} A(\bar B^0\to \pi^0\bar K^0) = -A(B^-\to \pi^-\bar K^0) + \sqrt{2} A(B^-\to \pi^0K^-)
-A(\bar B^0\to\pi^+K^-)$. Note that $\lambda_u^{(s)}/\lambda_s^{(s)}\sim \lambda^2\sim 0.04$ (with $\lambda$ the
Cabibbo parameter), so tree amplitudes are (at best) CKM suppressed. The full penguin amplitude
$\hat \alpha_4^p(\pi\bar K) = a_4^p(\pi\bar K) + r_\chi^K a_6(\pi\bar K)+\beta_3^p(\pi\bar K)$ contains a
scalar penguin amplitude $a_6^p$ and an annihilation amplitude $\beta_3^p$.
Both contributions are formally power corrections and will be discussed separately below.
The contributions from $Q_{1,2}$ to the leading penguin amplitudes $a_4^p$ have been recently calculated
at NNLO~\cite{1507.03700}:
\eqa{
a_4^u(\pi \bar{K})/10^{-2} &=& -2.87 -
[0.09 + 0.09i]_{\rm V_1} + [0.49-1.32i]_{\rm P_1} - [0.32+0.71i]_{\rm P_2}
\label{a4u} \\[2mm]
&&\hspace{-20mm} +\,\left[ \frac{r_{\rm sp}}{0.434} \right]
  \Big\{ 0.13 + [0.14 +0.12i]_{\rm HV} - [0.01-0.05i]_{\rm HP}
  + [0.07]_{\rm tw3} \Big\} = -2.46 - 1.94\,i\,,\nonumber\\[1em]
a_4^c(\pi \bar{K})/10^{-2} &=& -2.87 -
[0.09 + 0.09i]_{\rm V_1} + [0.05-0.62i]_{\rm P_1} - [0.77+0.50i]_{\rm P_2}
\label{a4c}\\[2mm]
&& \hspace{-20mm} +\,\left[ \frac{r_{\rm sp}}{0.434} \right]
  \Big\{ 0.13 + [0.14 +0.12i]_{\rm HV} + [0.01+0.03i]_{\rm HP}
  + [0.07]_{\rm tw3} \Big\} = -3.34 -1.05\,i\,. \nonumber
}
Spectator scatering (proportional to $r_{\rm sp}$) is numerically small.
The NNLO contribution is labeled `P$_2$', and it is found to be rather large and of the same order of the
NLO penguin contributions. It should be noted that in the case of $a_4^c$ there is a strong cancellation
at NLO (in the term labeled `P$_1$') between the two $Q_1$ contributions with different color topologies.
Thus the fact that the NNLO correction is much larger than the NLO seems accidental.
Stabilization of the $\mu$ dependence of the real parts suggests the perturbative expansion
is well behaved~\cite{1507.03700}.
Again, the scale dependence of the imaginary part is not significantly reduced at NNLO since the LO contribution
is real.

Full NNLO phenomenology for penguin decays would require the missing two-loop matrix elements of
penguin operators $Q_{3\cdots 6}$.

\subsection{Direct CP asymmetries}

Direct CP asymmetries require the interference of two amplitudes with different weak and strong phases.
Therefore they are governed by the penguin amplitude $\alpha_4^c$ and the imaginary parts in tree and
penguin amplitudes (strong phases). Since the leading-power leading-order amplitudes are real, strong phases
are either ${\cal O}(\alpha_s)$ or ${\cal O}(\Lambda/m_B)$. Since $\alpha_s/\pi \sim \Lambda/m_b$,
it is plausible that power corrections are ${\cal O}(1)$ effects in direct CP asymmetries. In addition,
perturbative corrections to CP asymmetries require tree and penguin amplitudes to NNLO, which is one
of the main motivations behind the calculations in Refs.~\cite{0911.3655,1507.03700}.

Direct CP asymmetries at NNLO for penguin decays have been discussed in~\cite{1507.03700}.
In Table~\ref{tab:acp} we reproduce some of the results for $B\to K \pi$ direct CP asymmetries.
In this case NNLO corrections are small because $a_4^p$ is only a part of the penguin amplitude $\hat \alpha_4^p$,
and $a_6^p$ is numerically large, thereby diluting the effect. The `NNLO' column does not include the
annihilation contribution $\beta_3^p$ nor the twist-3 spectator scattering contributions. These are
power suppressed but not calculable, and induce a significant error in the predictions. Using a similar
model for power suppressed non-factorisable contributions as in~\cite{0308039} these are included in
the column labeled `NNLO+LD', with the annihilation contribution $\beta_3^p$ giving the dominant effect.
In this case the agreement with data is improved, although uncertainties are inflated considerably.
The prediction and experimental number for the quantity
$\delta(\pi K) \equiv A_{\rm CP}(\pi^0K^-)-A_{\rm CP}(\pi^+K^-)$ are also given.
The theoretical error in this quantity is under better control because of certain cancellations
in hadronic uncertainties. The tension between theory and experiment in $\delta(\pi K)$ remains a ``puzzle''
(see \eg\cite{0707.0212}).
In the case of $PV$ and $VV$ final states such as $\rho K$, $\pi K^*$, $\rho K^*$, the NNLO contribution to CP
asymmetries can be significant, depending on the role of the scalar penguin amplitude $a_6^c$.
In any case it is a general feature that the long-distance annihilation contribution is very important
numerically. But experimental results for these $PV$ and $VV$ modes are still quite uncertain.

%%%%%%%%%%%%%%%%%%%%%%%%%%%%%%%%%%%%%%%%%%%%%%%%%%%%%%%%%%%%%%%%%%%

\begin{table}
\let\oldarraystretch=\arraystretch
\tabcolsep0.2cm
\renewcommand*{\arraystretch}{1.3}
\begin{center}
\begin{tabular}{lcccc}
\toprule[2pt]
%&&&&\\[-5mm]
$f$  & ${\rm NLO}$ & ${\rm NNLO}$ &  ${\rm NNLO}+{\rm LD}$ & Exp \\ \hline
%&&&&\\[-5mm]
%%
$\pi^-\bar{K}^0$
& $\phantom{-}0.71_{\,-0.14\,-0.19}^{\,+0.13\,+0.21}$
& $\phantom{-}0.77_{\,-0.15\,-0.22}^{\,+0.14\,+0.23}$
& $\phantom{-}0.10_{\,-0.02\,-0.27}^{\,+0.02\,+1.24}$
& $-1.7\pm1.6$ \\[0.5mm]
$\pi^0K^-$
& $\phantom{-}9.42_{\,-1.76\,-1.88}^{\,+1.77\,+1.87}$
& $10.18_{\,-1.90\,-2.62}^{\,+1.91\,+2.03}$
& $-1.17_{\,-0.22\,-\phantom{0}6.62}^{\,+0.22\,+20.00}$
& $\phantom{-}4.0\pm2.1$ \\[0.5mm]
$\pi^+K^-$
& $\phantom{-}7.25_{\,-1.36\,-2.58}^{\,+1.36\,+2.13}$
& $\phantom{-}8.08_{\,-1.51\,-2.65}^{\,+1.52\,+2.52}$
& $-3.23_{\,-0.61\,-\phantom{0}3.36}^{\,+0.61\,+19.17}$
& $-8.2\pm0.6$ \\[0.5mm]
$\pi^0\bar{K}^0$
& $-4.27_{\,-0.77\,-2.23}^{\,+0.83\,+1.48}$
& $-4.33_{\,-0.78\,-2.32}^{\,+0.84\,+3.29}$
& $-1.41_{\,-0.25\,-6.10}^{\,+0.27\,+5.54}$
& $\phantom{-1}1\pm10$ \\[0.5mm]
$\delta(\pi \bar{K})$
& $\phantom{-}2.17_{\,-0.40\,-0.74}^{\,+0.40\,+1.39}$
& $\phantom{-}2.10_{\,-0.39\,-2.86}^{\,+0.39\,+1.40}$
& $\phantom{-}2.07_{\,-0.39\,-4.55}^{\,+0.39\,+2.76}$
& $12.2\pm 2.2$ \\[0.5mm]
\bottomrule[2pt]
\end{tabular}
\end{center}
\caption{\label{tab:acp}
Direct CP asymmetries (in percent) for $\pi K$ final states (from Ref.~\cite{1507.03700}).}
\end{table}

%%%%%%%%%%%%%%%%%%%%%%%%%%%%%%%%%%%%%%%%%%%%%%%%%%%%%%%%%%%%%%%%%%%

Direct CP asymmetries will most certainly lead to a clear picture of successes and failures of leading-power
factorization. So far the situation is rather confusing, with an ``ununderstood pattern of agreements
and disagreements'' (quoting~\cite{1501.07374}). More precise data will also contribute to clarify the
situation, with good prospects from LHCb and \mbox{Belle-II}.

\subsection{Power corrections}

Power corrections are the main source or uncertainty in the prediction of non-leptonic two-body $B$-decay
amplitudes. Tests of leading-power factorization in $B\to D^{(\star)}L$ decays (with $L=\pi,\rho,K,K^\star$)
can be performed by considering ratios of non-leptonic to semileptonic $B\to D^{(\star)}\ell\nu$
or between different non-leptonic rates, where the factor $(V_{cb} \times {\rm form\ factor)}$
cancels~\cite{0006124}. NNLO theory predictions~\cite{1606.02888} for non-leptonic ratios agree within
uncertainties with current data, but predictions for non-leptonic-to-semileptonic ratios are
universally above the data by $~10-20\%$, with relatively low uncertainties~\cite{1606.02888}. A possible
interpretation is a universal (negative) power correction of $~10-15\%$ at the level of the amplitude, which would
give a positive contribution in non-leptonic-to-semileptonic ratios but would cancel out in purely non-leptonic
ones. This size of power corrections is at the level of expected ${\cal O}(\Lambda/m_b)$ contributions,
and does not invalidate the QCD-Factorization for heavy-light final states. While this gives also support
to the formalism in the case of charmless two-body decays, one cannot exclude the possibility
of enhanced power corrections in this case.

Some power corrections are calculable and numerically important.
This is the case of the scalar penguin amplitude
$a_6^p(M_1M_2)$, which contributes to the full penguin amplitude
$\hat \alpha_4^p(M_1M_2) = a_4^p(M_1M_2) \pm r_\chi^{M_2} a_6(M_1M_2)+\beta_3^p(M_1M_2)$. Here the plus (minus)
sign applies when $M_1$ is a pseudoscalar (vector) meson, and $r_\chi^{M_2}$ is a ``kinematic''
factor that contains a power suppression and a chiral enhancement,
\eg$r_\chi^{K} = 2 m_K^2/[m_b (m_q+m_s)]$. Numerically $r_\chi\simeq 1$,
so although the scalar penguin amplitude is power suppressed, it is numerically leading. This is not a problem
since this amplitude factorizes and it is therefore, calculable.

Other power corrections come from annihilation (\eg$\beta_3^p$) and spectator scattering contributions
--\eg{}the terms labeled `tw3' in Eqs.~(\ref{a1}),~(\ref{a2}),~(\ref{a4u}),~(\ref{a4c})--,
which do not factorize. As discussed above, annihilation
contributions are relevant for CP asymmetries. Modeling these power-suppressed contributions leads
to large uncertainties in the QCDF predictions.

One possibility is to parametrize the weak annihilation (WA) contributions and determine
whether some pattern for WA can accommodate the data. A global fit to most of the available data on
$B_{u,d,s}\to PP,\,VP,\,VV$ modes~\cite{1409.3252} finds that the SM can reproduce the experimental results
(with a few exceptions) using one universal WA parameter for each decay system, and with no anomalously
large values for these parameters (that is, consistent with the most popular model \eg\cite{0308039}).
The exceptions are $\delta(\pi K)$ (thus not resolving the ``$\Delta A_{CP}$ puzzle'') and, less significantly,
the branching ratio of $B^0\to K^{*0}\phi$, with a pull around $\sim 2\,\sigma$. Removing the
``universality'' assumption for WA will however ease all tensions (including $\Delta A_{CP}$), at the obvious cost
of more freedom and little predictivity. A similar analysis can be found in~\cite{1409.2995}.

Another possibility is to look for theoretical quantities where non-factorisable contributions cancel,
either completely or approximately. An example is given in Ref.~\cite{0603239}, where it is shown
how this cancellation takes place in the quantity $\Delta_f \equiv T^u_f-T^c_f$ [in the notation of Eq.~(\ref{amp})]
for certain penguin-mediated decays (for a list of such modes see~\cite{1111.4882}). Using the
QCDF prediction for this quantity one can predict certain relationships between observables
which can help to test branching ratios and direct CP asymmetries~\cite{0603239,0610109} or to extract
mixing angles from data~\cite{1111.4882,0705.0477}.

Now that perturbative calculations have reached the NNLO level, progress in the theoretical study of
non-leptonic two-body $B$ decay amplitudes requires addressing power corrections systematically.
This is strongly motivated given the experimental prospects for measurements of branching fractions and
CP asymmetries in two-body charmless $B$ decays.

\section{Three-body decays}
\label{three-body}

As in two-body decays, the challenge is to calculate the matrix elements
$\langle M_aM_bM_c | Q_i(\mu) | \bar B\rangle$ from first principles in QCD.
Alternatively (as in two-body decays) one may attempt to establish rigorous relationships between various of these matrix elements that can be exploited phenomenologically.
Either way, three-body decays are considerably more challenging than two-body decays, but they provide a series of theoretical and phenomenological advantages:
\begin{itemize}
\item[$\triangleright$] The number of different three-body final states is an order of magnitude larger than the number of
two-body decays. On top of that,  each final state depends on two kinematic variables, as opposed to two-body decays where the kinematics is fixed. This leads to a much
broader phenomenology.
\item[$\triangleright$] `Quasi-two-body' decays $\bar B \to M_a M (\to M_b M_c)$, where $M$ is a
strong resonance, are
only well defined in the context of the three-body decay.
Experimentally, quasi-two-body decays are extracted from the three-body
phase-space distribution by fitting to resonance models. Theoretically,
one resorts to the narrow-width approximation where $M$ is treated as a stable
particle.
A correct understanding of the three-body decay is necessary in order to compute corrections to the
quasi-two-body approximation.

\item[$\triangleright$] Three-body decays are used for spectroscopy, by looking for
resonant structures in 2-particle invariant-mass distributions (see \eg\cite{1605.03889,latham}).
The spin of such resonances can be determined as well by means of an angular analysis.

\item[$\triangleright$] Factorization properties of three-body decays will depend continuously on two kinematic variables. Thus three body decays are a tool for detailed data-driven tests of factorization and power corrections in $B$ decays.

\item[$\triangleright$] As discussed above, strong phases in two-body decays are either perturbative  --${\cal O}(\alpha_s(m_b))$--, or power suppressed --${\cal O}(\Lambda/m_b)$.
Direct CP asymmetries are then predicted to be correspondingly suppressed, and, since $\alpha_s(m_b)/\pi \sim \Lambda/m_b$, power corrections may be
numerically leading.
On the contrary, in three-body decays strong phases appear already
at the leading power and perturbative order, through complex phases in matrix elements such as
$F_\pi \sim \langle 0 | j | \pi\pi \rangle$ or $F^{B\pi\pi}\sim \langle \pi\pi | j | \bar B \rangle$.
While these phases may in principle not be calculable, these matrix elements can be obtained from data, from other \emph{unrelated} decay modes.
Local direct CP asymmetries can be large, with good prospects for model-independent predictions. This may lead to improved extractions of CKM angles
from direct CP violation.
\end{itemize}

It is fair to say that the theory of three-body non-leptonic decays is still in an early stage of development.
Much attention has been put on three-body decays as a tool to study two-meson systems (see \eg\cite{1605.03889}).
Recent work also includes model studies of 2- and 3-body final-state interactions, and implications of
flavor symmetries. We will briefly review a few of these ideas, before focusing on the approaches based
on factorization.

\subsection{Kinematics}

\begin{figure}[t]
\includegraphics[width=7.2cm]{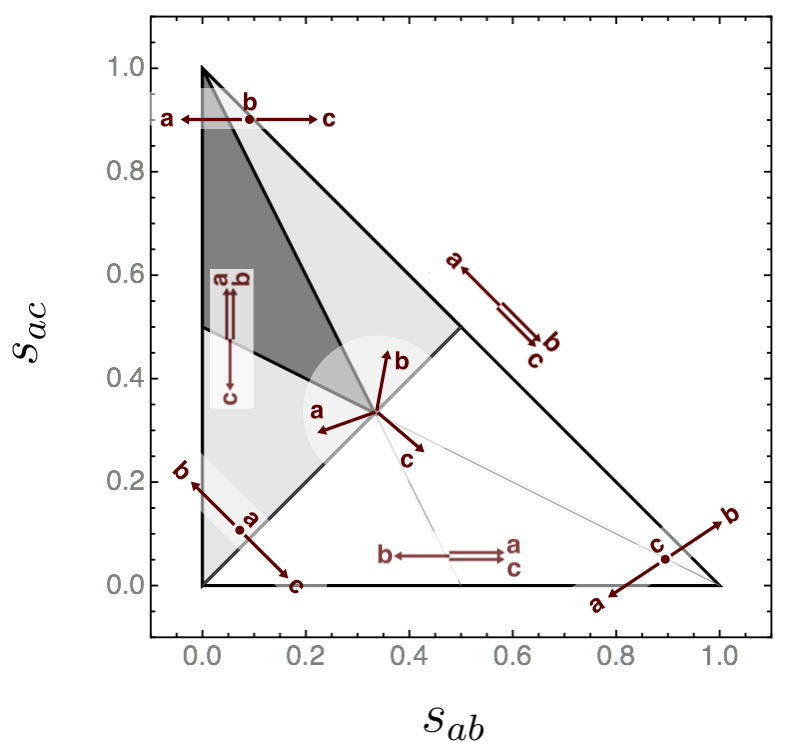}
\hspace{4mm}
\includegraphics[width=7.6cm]{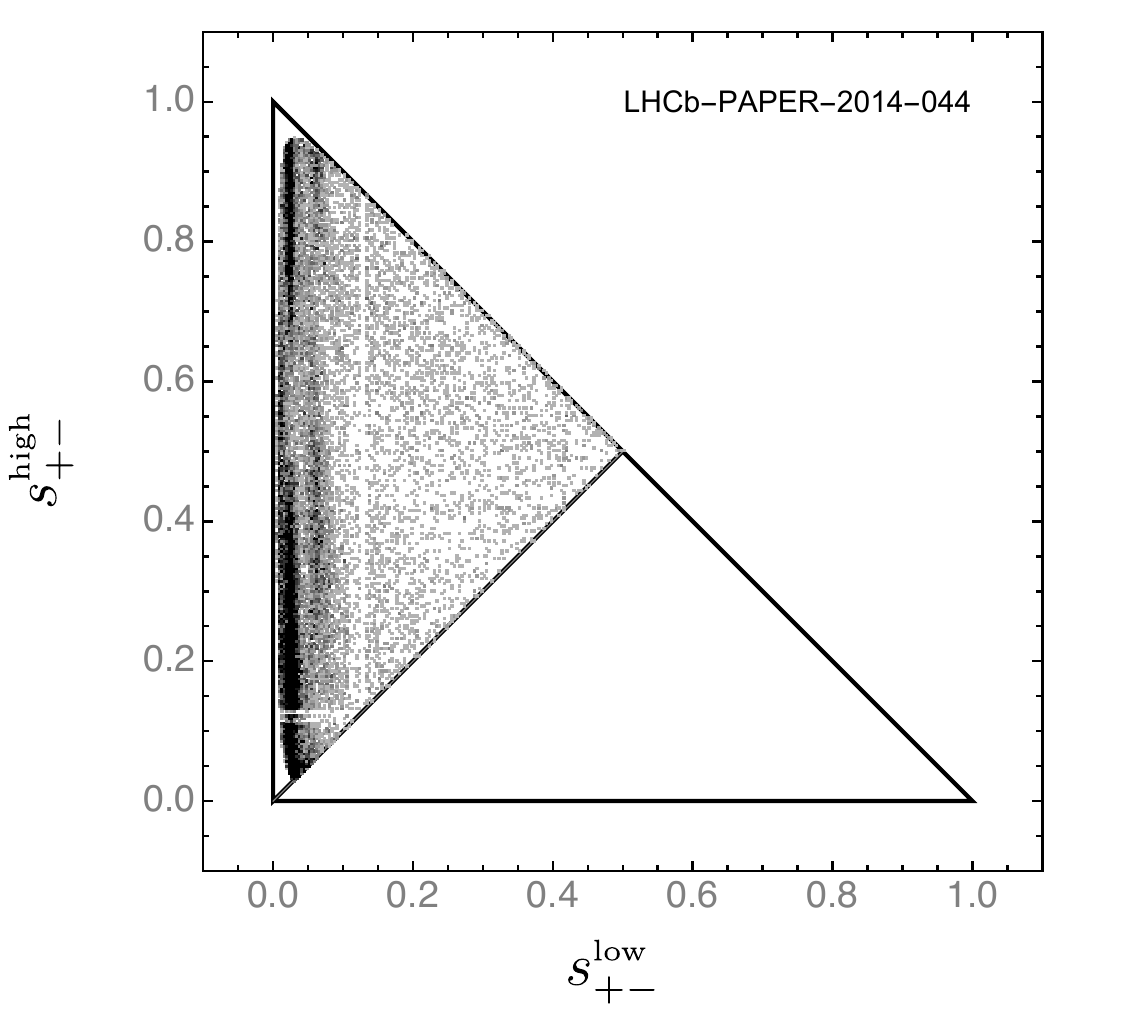}
\caption{Phase space of the three-body decay $B\to M_a M_b M_c$ in terms of the normalized invariants $s_{ab},s_{ac}$.
Left plot: Special kinematic configurations are indicated. If $M_b=M_c$ or $M_a=M_b=M_c$ then $s_{ab}\to s_{ab}^{\rm low}$ and
$s_{ac}\to s_{ab}^{\rm high}$, and the phase space is reduced to the light-gray region and dark-gray region respectively.
Right~plot: An example of the $M_b=M_c$ case: $B^- \to \pi^+\pi^-\pi^-$ Dalitz distribution from LHCb~\cite{1408.5373}.}
\label{DPs}
\end{figure}

While the kinematics of two-body decays is fixed, three-body decay amplitudes depend
on two kinematic variables.
We define the kinematics of the three-body decay by
\eq{
\bar B (p_B) \to M_a(p_1) M_b(p_2)M_c(p_3).
}
It is customary to take these variables as two invariant masses
of two pairs of final state particles:
\eq{
s_{ab} = (p_1 + p_2)^2/m_B^2\ ,\quad s_{ac} = (p_1 + p_3)^2/m_B^2\ ,
}
where we normalize by $m_B^2$ for convenience.
All physical kinematic configurations thus define a two-dimensional region in the $s_{ab}$-$s_{ac}$ plane,
which in the limit where all final particles are massless is a triangle defined by
$s_{ab}>0$, $s_{ac}>0$, $s_{ab}+s_{ac}<1$. (We will assume massless decay products in the following
for simplicity.) In the case of having identical particles in the final state, we can label these by their
momenta, removing thus any ambiguity. This reduces the phase space in half (in case of two identical particles)
or one-sixth (in the fully symmetric case), see~Fig.~\ref{DPs}.

The amplitude of the three-body decay is a function of the two kinematic variables, ${\pazocal A}(s_{ab},s_{ac})$.
The differential decay rate is given by:
\eq{
\frac{d^2\Gamma}{ds_{ab}\,ds_{ac}}=\frac{m_B}{32(2\pi)^3} |{\pazocal A}(s_{ab},s_{ac})|^2
}
and the corresponding distribution in the phase-space region is called a Dalitz plot.

The Dalitz plot can be divided in different regions with ``characteristic'' kinematics.
We refer always to the $B$-meson rest frame.
The central region corresponds to the case where all three final particles are ejected with large energy
in a ``mercedes star''-like configuration.
In this case all invariant masses are large.
The corners correspond to the situation in which one final particle is approximately at
rest, and the other two fly back-to-back with large energy. In this case one invariant mass is large, and the
other two are small.
At the central part of the edges the kinematics is such that two particles move collinearly with large energy
and the other particle recoils back. In this case one invariant mass is small and the other two are large. These regions with characteristic kinematics are sketched in Fig.~\ref{DPs}.

\subsubsection{Partial-wave expansions and isobar model}

The Dalitz plot distribution is presumably dominated by resonant quasi-two-body configurations along its edges.
For example, in the right plot in Fig.~\ref{DPs} we see that the event distribution in $B^-\to \pi^-\pi^+\pi^-$
shows a concentration along the region of small $s_{+-}$, corresponding to the quasi-two-body decay
$B^-\to \pi^- \rho^0$, and possibly also $B^-\to \pi^- R$ with $R=f_0(980),\rho'(1450)$, etc.
No such concentration is seen along the $s_{--}$ edge, since there are no relevant resonances in that channel.
To a first approximation, one can then describe the three-body decay as a coherent sum of
quasi-two-body decays $\bar B \to R^{(\ell)}_{ij}(\to M_i M_j) M_k$, with $R^{(\ell)}_{ij}$
denoting a spin-$\ell$ resonance in the $(ij)$ channel. This resonance contributes to the region
$s_{ij}\sim (m_{R_{ij}}\pm \Gamma_{R_{ij}})^2/m_B^2$, where $m_{R_{ij}}, \Gamma_{R_{ij}}$ are its mass and
width respectively. The profile of this contribution in the other variable $s_{ik}$ depends on the spin of the
resonance. It is thus convenient to expand the amplitude in partial waves in the corresponding channel.
For example, when considering resonances in the $(ab)$ channel, one may trade the variable $s_{ac}$ by the angle
$\theta_{c}$ between the momenta $\vec p_1$ and $\vec p_3$ in the $(M_aM_b)$ rest frame. In the massless limit this
angle is given by
\eq{
\cos\theta_c = \frac{1 - s_{ab} - 2 s_{ac}}{1-s_{ab}} \ .
}
The amplitude ${\pazocal A}(s_{ab},s_{ac})$ can then be expressed in terms of $(s_{ab},\theta_c)$ and expanded
in Legendre polynomials:
\begin{equation}
{\pazocal A}(s_{ab},s_{ac}) = \sum_{\ell=0}^\infty (2\ell+1) \, {\pazocal A}^{(\ell)}(s_{ab})
\,P_\ell(\cos\theta_c)\ .
\label{eq:3-bodyPWE}
\end{equation}
Resonances of spin=$\ell$ in the $(ab)$ channel will contribute only to the corresponding partial wave.
The partial wave expansion is only useful if it can be truncated. But such a truncation makes the
r.h.s of~(\ref{eq:3-bodyPWE}) algebraic in $s_{ac}$, so the presence of physical singularities of
${\pazocal A}(s_{ab},s_{ac})$ in the $(ac)$ channel imply that the partial wave expansion cannot
converge~\cite{1409.8652}.
A solution to this problem is provided by the \emph{isobar model},
where the amplitude is written as a sum of a finite number of partial waves in all three channels:
\eq{
{\pazocal A}(s_{ab},s_{ac}) = \sum_{\ell=0}^{\ell_{\rm max}} (2\ell+1)\, a_{\ell}^{ab}(s_{ab}) P_\ell(\cos\theta_c)
\ \ +\ \ (abc\to bca) \ \ + \ \ (abc\to cab)\ .
\label{Isobar}
}
The isobaric amplitudes $a_\ell(s)$ are typically modeled by energy-dependent Breit-Wigner amplitudes,
although more sophisticated line-shapes are also common, depending on the details of the spectrum.
One can see in the right plot in Fig.~\ref{DPs} that there is also an approximately constant background
in the $B^-\to \pi^-\pi^+\pi^-$ Dalitz distribution.
To account for such a ``non-resonant'' background, a smooth component may be added to the amplitude
${\pazocal A}(s_{ab},s_{ac})$ in Eq.~(\ref{Isobar}), but the exact kinematic dependence
of this component is rather arbitrary.

\subsubsection{Final-state interactions and CPT constraint}

Isobaric amplitudes may contain strong phases (for example of the Breit-Wigner type), and the complicated
interference between such amplitudes in all channels may lead to large and complicated localized CP asymmetries in
the model for the three-body decay. But the isobar model does not account for
coupled-channel effects (beyond resonance interference) or three-body rescattering. These type of
final-state interactions are often invoked as yet another source of non-perturbative strong phases
which may be relevant to describe CP violation across the Dalitz plane.
These effects can be analyzed by means of dispersive methods, of may be modeled
separately~\cite{1409.8652,1508.06841,1512.09284,1609.01568}.

One additional constraint may be obtained by combining unitarity and CPT invariance of strong and electroweak
interactions separately.
This implies that, not only the total rates of particles and antiparticles are equal,
but also that partial rates into same-flavor states are equal~\cite{Wolfenstein:1990ks}:
\begin{equation}
\sum_f \ \big[\Gamma(B\to f) -\Gamma(\bar B\to \bar f) \big] = 0\ ,
\label{eq:3-bodyCPT}
\end{equation}
where the sum runs over all final states $f$ with the same flavor quantum numbers.
The individual exclusive rates may not be equal, as there might be direct CP violation
in the exclusive modes, but all CP asymmetries of exclusive decays to same-flavor final
states must add up to zero, leading to a compensation mechanism.
It is clear how this constraint is useful in cases where a very limited of final states is available,
but in the case of $B$-meson decays the constraint~(\ref{eq:3-bodyCPT})
is by itself of little use. This constraint, however, may be imposed on very simple models
which consider a just a few coupled channels. This may provide some insight on the pattern
of final-state interactions, and the pattern of CP asymmetries in different modes.
An an example, a model with two coupled S-wave
$(\pi^+\pi^-)$ and $(K^+K^-)$ states consistent with the CPT constraint~\cite{1307.8164}
provides a qualitative understanding of the observed CP asymmetries in $B^\pm\to K^\pm \pi^+\pi^-$
and $B^\pm\to K^\pm K^+K^-$ in the region $1\,{\rm GeV}^2\lesssim m_{\pi\pi}^2,m_{KK}^2 \lesssim 2.2\,{\rm GeV}^2$,
where S-wave $\pi^+\pi^- \leftrightarrow K^+K^-$ scattering is supposed to be dominant.
In this case the CPT constraint seems to explain why these asymmetries (when properly weighted by the branching ratios)
are approximately equal and of opposite sign. A similar pattern is observed in
$B^\pm\to \pi^\pm \pi^+\pi^-$ and $B^\pm\to \pi^\pm K^+K^-$. More complicated models which include
$\rho(770)$ and $f_0(980)$ resonances have also been studied in this framework~\cite{1506.08332}.

\subsubsection{Naive factorization}

In the case of two-body charmless $B$ decays, naive factorization arises in the heavy-quark limit and
at the leading order in $\alpha_s(m_b)$~\cite{9905312}, and it is in that sense a prediction of QCD.
Perturbative ``non-factorisable'' corrections can be computed in the framework of QCD-factorization.
Such a theory has not been fully developed in the three-body case, but many phenomenological analyses exist
which \emph{assume} that ``naive factorization plus $\alpha_s$ corrections'' is also a valid approach to
three-body decays.
This might indeed be the case in the kinematic regions around the edges of the Dalitz plot, which
contain ``quasi-two-body'' configurations such as $B\to\rho\pi$, to which the two-body QCD-factorization formula
applies~\cite{0308039}.

Consider the decay $\bar B\to M_aM_bM_c$ in the kinematic region where $s_{bc}\ll 1$.
We denote the two-meson system with small invariant mass by $[M_bM_c]$.
The naive factorization formula for the amplitude $A_{M_a[M_bM_c]}^p = T_{M_a[M_bM_c]}^p + P_{M_a[M_bM_c]}$ [in
the notation of Eq.~(\ref{amp})]
is given by~\cite{Dedonder:2010fg}:
\eq{
A_{M_a[M_bM_c]}^p = \sum_k \bigg[ a_k^p(M_a,[M_bM_c]) \,A^k_{M_a,[M_bM_c]}
+ a_k^p([M_bM_c],M_a) \,A^k_{[M_bM_c],M_a}  \bigg]
}
where
\eq{
A^k_{M_a,[M_bM_c]} = -\frac{4 G_F}{\sqrt2} \langle  M_a| j_k^1|\bar B \rangle \langle [M_bM_c]| j_k^2 |0 \rangle \ ,\quad
A^k_{[M_bM_c],M_a} = -\frac{4 G_F}{\sqrt2} \langle  [M_bM_c]| j_k^1|\bar B \rangle \langle M_a| j_k^2 |0 \rangle \ .\quad \nonumber
}
Here $j_k^{1,2}$ are local bilinear color-singlet currents related to the operators in the effective Lagrangian,
and $a_k^p$ are the usual coefficients in QCDF~\cite{0308039}.
Annihilation contributions and hard-scattering corrections are typically neglected in these analyses.
NLO vertex corrections and penguin contractions in $a_k^p(M_a,[M_bM_c])$ would involve a light-cone distribution
amplitude of the pair $[M_bM_c]$ (see~\cite{0308039}). A simple way to implement these corrections here is to adopt
a multi-resonance model~\cite{Dedonder:2010fg}.
This resonance model is also used to calculate the matrix elements $\langle  [M_bM_c]| j_k^1|\bar B \rangle$
and $\langle [M_bM_c]| j_k^2 |0 \rangle$.
Note that this requires a partial wave decomposition in the $(bc)$ channel,
which (as discussed above) involves immediately all values of $s_{ab}$, including the corners of the Dalitz plot
(where $M_b$ or $M_c$ are soft). One must then reconsider whether the naive factorization ansatz is
expected to hold when two invariant masses are small (and not only one). Note also that in that case one can
write a different factorized form of the amplitude: if $M_b$ is soft then the amplitudes $A_{M_a[M_bM_c]}$
and $A_{[M_aM_b]M_c}$ must coincide at the corner of the Dalitz plot.

A more adventurous approach is to extend this factorization formula to the whole Dalitz
plot~\cite{Cheng:2007si,Cheng:2016shb}. In this way one obtains a more complete set of predictions,
while giving up the previous theoretical arguments for factorization.
These phenomenological analyses include in addition an estimate of non-resonant contributions in the following way:
The $B\to M_bM_c$ form factors $\langle  [M_bM_c]| j_k^1|\bar B \rangle$ are calculated in the framework of the
heavy-meson chiral perturbation theory (\mbox{HM$\chi$PT}). The HM$\chi$PT applies in the kinematic region where the
two mesons are soft. This region is unphysical in charmless three-body decays.
Using an exponential one-parameter ansatz, the result is extrapolated to the whole physical region.
This parameter is assumed universal, and is fitted to the ``non-resonant" component
of $B^- \to \pi^-\pi^+\pi^-$ provided by the $B$-factories. The resulting value is used to predict non-resonant
contributions in other modes~\cite{Cheng:2007si,Cheng:2016shb}.
The (model-dependent) predictions obtained within this ``extended factorization'' approach are in fair agreement
with data for $B^- \to K^- K^+ K^-$ and $B^- \to K^- \pi^+ \pi^-$ branching fractions~\cite{Cheng:2016shb},
but the significance of this agreement is not easy to interpret. CP violation is also predicted,
giving the right pattern for some modes, but not for others~\cite{Cheng:2016shb} (for example the resulting CP asymmetries
for $B^-\to \pi^- K^+K^-$ and $B^-\to K^- \pi^+\pi^-$ are found to have the wrong sign;
in this case, the CPT constraint discussed in the previous section may help to understand the problem).

\subsection{Three-body decays in QCDF}

The two kinematic invariants on which the decay amplitudes depend, introduce two extra scales in the problem.
Different forms of factorization theorems may apply in different regions of the Dalitz plot depending on the scaling
of these momentum scales with $m_b$~\cite{talks}. We will restrict ourselves hereon to the example of
$\Bar B \to \pi\pi\pi$ for simplicity.

In the central region, where all invariant masses are of order $m_B$
($s_{12}\sim s_{13}\sim 1/3$), the following formula is conjectured~\cite{1505.04111}:
\begin{equation} 
\langle \pi\pi\pi | Q_i |\bar B \rangle_{s_{ij} \sim 1/3} = F^{B\to \pi} \, T_i^I \star \Phi_{\pi} \star \Phi_{\pi}
+ T_i^{II} \star \Phi_B \star \Phi_{\pi} \star \Phi_{\pi} \star \Phi_{\pi} \ ,
\label{eq:FactCenter}
\end{equation} 
where we have written the convolutions of hard-scattering kernels and distribution amplitudes
schematically. The kernels $T_i^{I,II}$ can be computed perturbatively in QCD, and are related
to matching coefficients of SCET operators such as
\begin{equation}
\big[ (\bar\xi_1 W_{1})(t n_1)\Gamma_i^1(W_{1}^\dagger \xi_1)(0) \big]
\big[ (\bar\xi_2 W_{2})(t n_2)\Gamma_i^2(W_{2}^\dagger \xi_2)(0) \big]
\big[ (\bar\xi_3 W_{3})(0) \Gamma_i^3 h_v(0)\big]
\end{equation}
with fields in three collinear directions $\{n_1,n_2,n_3\}$, generalizing the discussion in Section~\ref{scfact}.
The three collinear directions satisfy $n_i\cdot n_j \gg \Lambda_{QCD}/m_b$ (for $i\ne j$), such that a gluon
coupling to two different collinear modes is offshell by a large amount. 

\begin{figure}[t]
\raisebox{6mm}{\includegraphics[width=70mm]{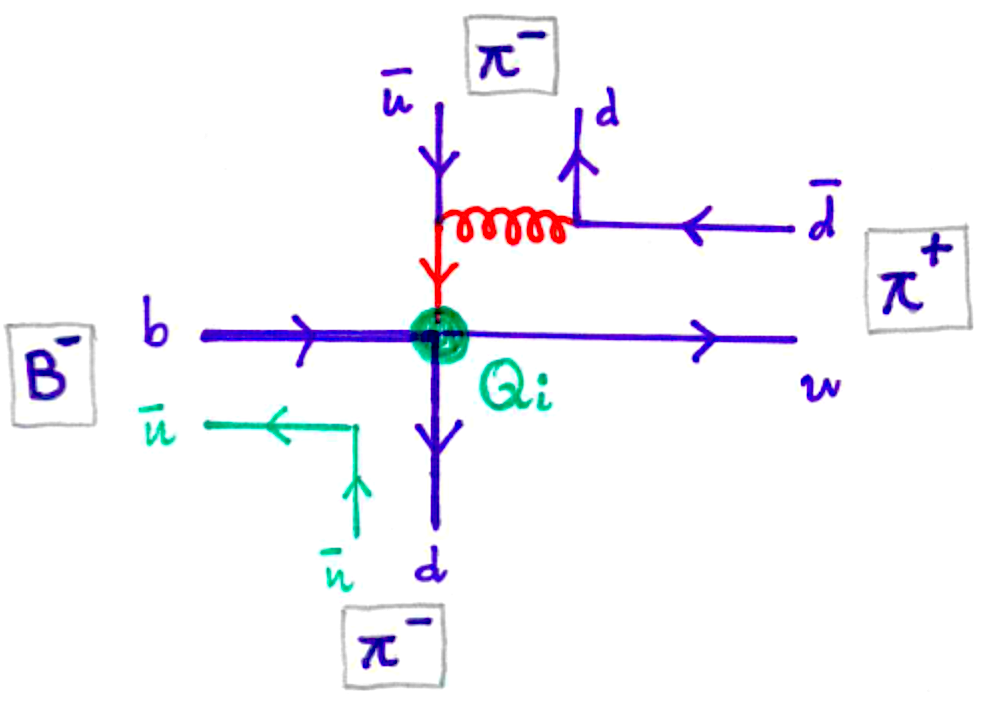}}
\hspace{10mm}
\includegraphics[width=65mm]{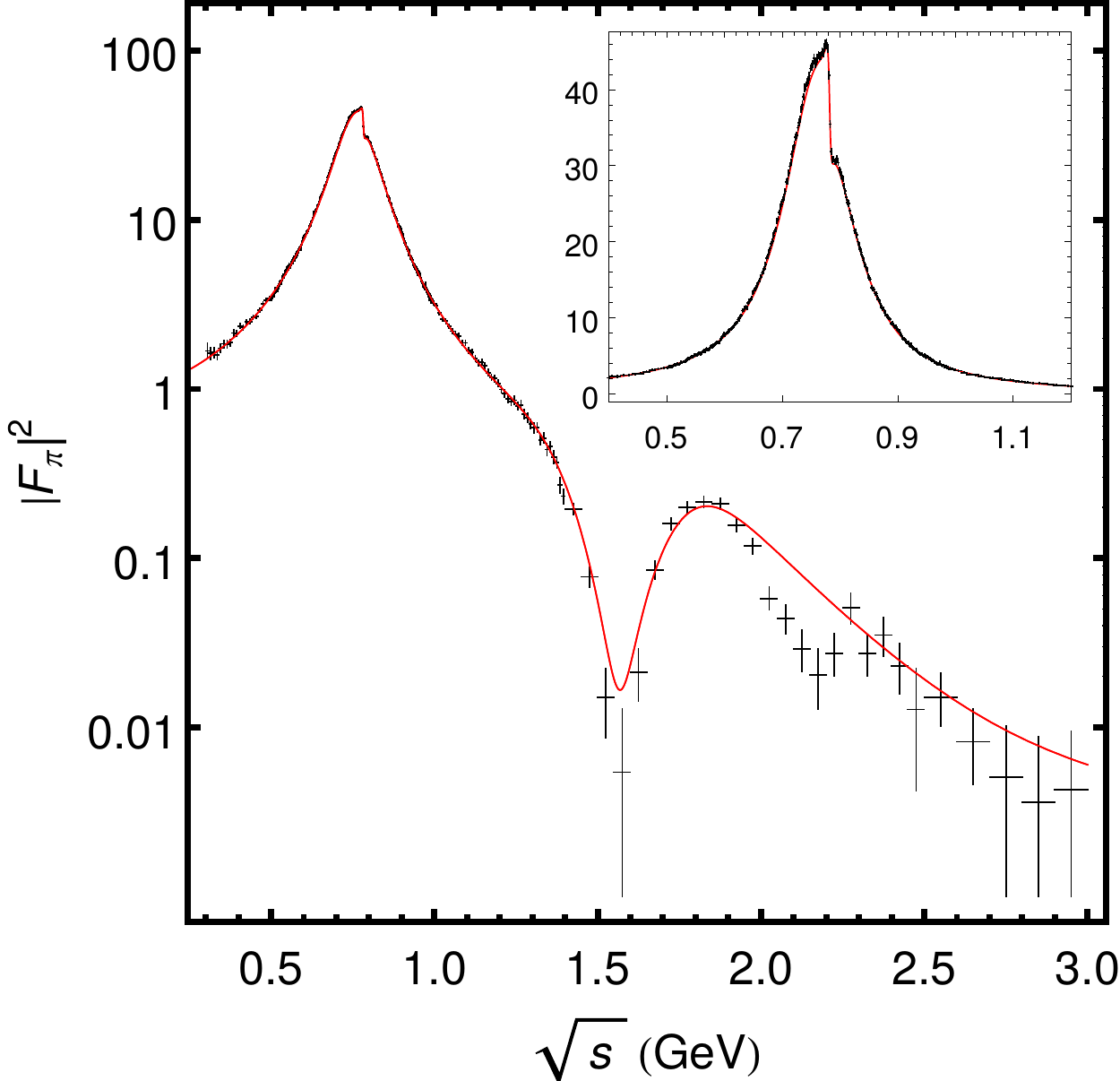}
\caption{
Left: Sample diagram contributing to $T_i^I$ in Eq.~(\protect\ref{eq:FactCenter}). The red propagators have a large
virtuality of order ${\cal O}(m_b)$, but become soft when $m_{\pi^+\pi^-}^2\sim \Lambda_{QCD}^2$ (towards the edge).
Right: Magnitude of the pion vector form factor $F_{\pi}(s)$ obtained by Babar form from
$e^+ e^- \to \pi^+\pi^- (\gamma)$~\cite{babarFpi}.
}
\label{fig2} 
\end{figure}

In this case, $T_i^I$ starts at ${\cal O}(\alpha_s)$ and $T_i^I$ starts at ${\cal O}(\alpha_s^2)$.
We consider only $T_i^I$, arising from
diagrams with an insertion of the operator $Q_i$ and all possible insertions of a hard
gluon splitting into a quark-antiquark pair with large invariant mass (see left panel of Fig~\ref{fig2}).
The convolutions of the resulting hard kernels $T_i^I$ with the pion LCDAs can be computed without
encountering end-point divergences, thus providing a check of the factorization formula to ${\cal O}(\alpha_s)$.
This is a non-trivial check because the kernels $T_i^I(u,v)$ already depend on the momentum fraction of the quarks
at the leading order (contrary to the two-body case), so the convolutions are non-trivial.

At the edges of the Dalitz plot, one invariant mass becomes small, and low energy
interactions between the corresponding pair of final state particles leads eventually
to the formation of resonances. This is the case for \eg$B\to\pi^+\pi^-\pi^+$ in the region
where $m_{\pi^+\pi^-} \sim m_{\rho}$, and appears as a band in the Dalitz plot.
The decay thus looks very much like a two-body decay, and one expects a similar
factorization formula, except for the fact that one particle is, instead, two~\cite{1505.04111}:
\begin{eqnarray}
\langle \pi^a\pi^b\pi^c|Q_i|B\rangle_{s_{bc} \ll 1} &=& F^{B\to \pi^a}\ T_a^I \otimes \Phi_{\pi^b\pi^c}
+ F^{B\to \pi^b\pi^c}\ T_{bc}^I \otimes \Phi_{\pi^a}
+ T^{II} \otimes \Phi_B \otimes \Phi_{\pi^a} \otimes \Phi_{\pi^b\pi^c}\ .\qquad
\label{edge}
\end{eqnarray}
Here $\Phi_{\pi\pi}$ denotes a two-pion distribution amplitude (2$\pi$LCDA), and $F^{B\to \pi\pi}$
denotes a $B\to\pi\pi$ form factor. This factorization formula is theoretically at the same level of
rigor as the one for two-body decays into unstable particles (\eg$B\to\rho\pi$), but
requires a more complicated hadronic input (see the following section).
This complication is the only cost of going beyond the narrow-width approximation and including non-resonant effects in
quasi-two-body decays.

The three-body amplitude at the central region --Eq.~(\ref{eq:FactCenter})-- is both power-suppressed and
$\alpha_s$-suppressed with respect to the amplitude at the edge, Eq.~(\ref{edge}).
There are certain parts of the central region amplitude that arise from factorization of
2$\pi$LCDAs or $B\to\pi\pi$ form factors at large dipion masses,
and the correspondence of such parts of the amplitudes can be checked analytically~\cite{1505.04111}.
This provides checks of the calculation, but also serves to understand the interpolation between the two regions.
Numerically, one finds that the part of the amplitude at the central region corresponding
to the large dipion limit of the 2$\pi$LCDA part of the amplitude at the edge agrees well
with the latter only for $m_B\gtrsim 20$~GeV, but not for realistic values, suggesting that
power corrections to Eq.~(\ref{eq:FactCenter}) are too large in reality, precluding a description
of the central region in terms of single pion states~\cite{1505.04111}. This is based on model extrapolations
of the pion vector form factor to larger energies; further study of two-pion states at higher invariant masses
would be desirable.

\subsection{Generalized hadronic input}

\subsubsection{Two-pion light-cone distributions}

The relevant two-pion distribution amplitude in Eq.~(\ref{edge}) is given by the following non-local matrix
element~\cite{1505.04111,9809483}
\eq{
\Phi_{\pi\pi}^q(z,\zeta,k_{12}^2) = \int \frac{dx^-}{2\pi} e^{i z (k_{12}^+ x^-)}
\langle \pi^+(k_1)\pi^-(k_2)| \bar q (x^- n_-) W_x \,\slashed{n}_+ q(0) |0\rangle\ ,
}
where $k_{12}^\mu=k_1^\mu+k_2^\mu\simeq (k_{12}^+/2) n_+^\mu$, $\zeta = k_{12}^+/k_1^+$, and $W_x$ is
a Wilson line which ensures that the non-local current is gauge invariant. At the leading order in $\alpha_s$,
the kernel $T_a^I$ in Eq.~(\ref{edge}) does not depend on $z$, and therefore we only need the normalization for
$\Phi_{\pi\pi}$:
\eq{
\int dz\,\Phi^q_{\pi\pi}(z,\zeta,s) = (2\zeta -1) F_\pi(s)
}
where $F_\pi(s)$ is the pion vector form factor. The magnitude of $F_\pi$ is well
known experimentally up to $s\sim 7$~GeV$^2$ (see right panel of Fig.~\ref{fig2}).
Higher moments of $\Phi_{\pi\pi}$ would be needed at higher orders, but these are much less known.

\subsubsection{$B\to \pi\pi$ form factors}

$B\to\pi\pi$ form factors can be obtained from $B\to\pi\pi\ell\nu$~\cite{1310.6660}.
The Lorentz structure of the leading-order $B^-\to\pi^-\pi^+\pi^-$ amplitude at low $m_{\pi^+\pi^-}$
is such that the relevant $B\to \pi\pi$ form factor is
\eq{
F_t(\zeta,k_{12}^2) \equiv -\frac1{\sqrt{q^2}}
\langle \pi^+(k_1)\pi^-(k_2)| \bar u\, \slashed{q}\,\gamma_5\, b |B^-(p)\rangle
}
where $q=p-k_{12}$ (in our case $q^2=m_\pi^2$).
At low dipion masses, this form factor may be studied by means of light-cone sum rules.
One may consider light-cone sum rules with two-pion distribution amplitudes~\cite{1511.02509}
or with $B$-meson distribution amplitudes~\cite{inprep}. In the first case one arrives
to a closed expression for $F_t$ in terms of moments of the 2$\pi$LCDA~\cite{alexWorkshop}:
\eq{
F_t (k_{12}^2,\zeta) =
\frac{m_b^2 \, m_\pi}{\sqrt{2} f_B m_B^2}
\int_{u_0}^1 \frac{du}{u^2} (m_b^2  + u^2 k_{12}^2)\ 
\Phi_{\pi\pi}^q(u,\zeta,k_{12}^2)\ 
e^{\frac{m_b^2}{M^2} - \frac{m_b^2  + u\,\bar u\, k_{12}^2}{u M^2}}\ .
}
Unfortunately, higher moments of the 2$\pi$LCDA are not known, and further study is required to extract the full power
of this sum rule.

In the second case, one starts with a non-local correlator between the $\bar B$ meson and the vacuum,
obtaining a sum-rule that depends on a convolution of the $B\to\pi\pi$ form factor and the pion vector
form factor~\cite{inprep}:
\eqa{
&&\hspace{-10mm} \int_{4m_{\pi}^2}^{s_0^{2\pi}} ds~e^{-s/M^2}~
\frac{s\ \sqrt{q^2}\ [\beta_{\pi}(s)]^2 }{4 \sqrt{6} \pi^2 \sqrt{\lambda}}
\, F^\star_{\pi}(s)\,F_{t}^{(1)}(s)
\\[2mm]
&&
\hspace{-9mm} 
=- f_B m_B^2 m_b\ \Bigg\{ \int_0^{\sigma^{2\pi}_0} d\sigma~e^{-s (\sigma,m_\pi^2)/M^2} 
\bigg[\frac{\sigma}{\bar{\sigma}} \phi_-^B(\sigma m_B)
-\frac{1}{\bar{\sigma}m_B}\bar{\Phi}_{\pm}^B(\sigma m_B) \bigg]
+ \Delta A_0^{BV}(m_\pi^2,\sigma_0^{2\pi},M^2)
\Bigg\} \ .\nonumber
}
where $F_t^{(1)}$ is the $P$-wave form factor, and $\Delta A_0$ denotes 3-particle contributions.
This sum rule depends on the $B$-meson LCDAs and not on the 2$\pi$LCDA.
While it does not provide the form factor in a closed form, this sum rule allows to test models for the $B\to\pi\pi$ form
factor,  and in the limit where the pion form factor is dominated by a zero-width $\rho$ meson,
one recovers analytically the well-known sum-rule for the $B\to\rho$ form factor $A_0^{B\rho}$~\cite{0611193}.

The $B\to\pi\pi$ form factors can also be calculated in the kinematic region where both pions are soft,
using a combination of dispersion theory and HM$\chi$PT~\cite{1312.1193}. This kinematic region is not directly
accessible from charmless three-body $B$ decays.
At large dipion masses, a factorization formula for the $B\to \pi\pi$ form factors has also been
proven at NLO recently~\cite{1608.07127}. As discussed above, this also proves part of the factorization formula
in Eq.~(\ref{eq:FactCenter}) at NLO.

\section{Conclusions and future prospects}

QCD Factorization is by now very well established as a QCD-based approach to charmless non-leptonic two-body decays. Perturbative calculations of hard-scattering
kernels have reached the NNLO precision, proving factorization to two loops and
confirming a good behavior of the perturbative expansion.

The pattern of branching
fractions is understood qualitatively, although some tensions are observed, mostly
in modes dominated by the color-suppressed tree amplitude. These tensions could be related to the spectator scattering contribution, which is proportional
to $\lambda_B$, the inverse moment of the $B$-meson LCDA, and which is currently not very well known. Values of $\lambda_B\sim 200$~MeV are favored, much lower than
sum-rule estimates. A direct experimental determination of $\lambda_B$ must await to 
a precise measurement of $B\to \gamma\ell\nu$ at Belle-II.

On the other hand, the recent calculation of penguin amplitudes
at NNLO provides the first perturbative corrections to CP asymmetries.
However, in this case power corrections could be ${\cal O}(1)$ effects,
explaining why the global picture in the comparison of theory and experiment is
far from clear. In addition, the ``$\Delta A_{\rm CP}$ puzzle" remains.
One should add that experimental measurements of CP asymmetries
to $PV$ and $VV$ final states are still not very precise.

Power corrections is now most probably the most pressing issue in order to make
progress in the theoretical understanding of charmless two-body decays, but the prospects are rather modest.

Three-body decays remain mostly unexplored from the theoretical point of view, 
although detailed and exciting experimental analyses of branching fractions and CP violation are piling up. We also expect many results from Belle-II. Recent studies
pursuing factorization methods for three-body decays look promising.

\section*{Acknowledgements}
I would like to thank David Hitlin and the organizers of FP\emph{CP}2016 for inviting me to
give this review talk and for a very enjoyable conference. I also thank Thomas Mannel and Susanne Kr\"ankl
for recent collaboration on three-body non-leptonic decays, and Tobias Huber for discussions and
for valuable comments on this draft.
I also thank all the participants of the worskhop {\it Future Challenges in Non-Leptonic B Decays: Theory and Experiment},
for making it super successful.
My research is currently funded by the Swiss National Science Foundation.

\end{document}